\documentclass[final,3p,times,twocolumn,sort&compress]{elsarticle}
\journal{Nuclear Physics B}
\usepackage{fullpage}
\usepackage[centertags]{amsmath}
\numberwithin{equation}{section}
\usepackage{amssymb}
\usepackage{bm}
\usepackage{graphicx}
\usepackage{multirow}
\usepackage{mathtools}
\usepackage{url}
\usepackage[hyperindex]{hyperref}
\usepackage{subfigure}
\usepackage{stackrel}
\usepackage{xspace}
\newcommand{\nua}[1]{\ensuremath{\rlap{\kern-1.0pt\ensuremath{\overset{\scriptscriptstyle(-)}{\phantom{\nu}}}}{\ensuremath{{\nu}_{#1}}}}\xspace}

\begin{document}

\begin{frontmatter}

\title{\textbf{Light Sterile Neutrinos: Status and Perspectives}\tnoteref{titlab}}
\tnotetext[titlab]{Invited contribution to the Nuclear Physics B Special Issue on Neutrino Oscillations celebrating the Nobel Prize in Physics 2015.}

\author{Carlo Giunti}

\address{INFN, Sezione di Torino, Via P. Giuria 1, I--10125 Torino, Italy}

\begin{abstract}
The indications in favor of the existence of light sterile neutrinos
at the eV scale found in short-baseline neutrino oscillation experiments is reviewed.
The future perspectives of short-baseline neutrino oscillation experiments
and the connections with $\beta$-decay measurements of the neutrino masses
and
with neutrinoless double-$\beta$ decay experiments are discussed.
\end{abstract}


\end{frontmatter}

\section{Introduction}
\label{sec:intro}

The 2015 Nobel Prizes in Physics
is a great acknowledgment of the fundamental importance of
the model-independent discoveries of neutrino oscillations
in the Super-Kamiokande atmospheric neutrino experiment
\cite{Fukuda:1998mi}
and in the SNO solar neutrino experiment
\cite{Ahmad:2002jz}.
These discoveries, which proved that neutrinos are massive and mixed particles,
led to the standard three-neutrino mixing paradigm ($3\nu$),
in which the three active neutrinos
$\nu_{e}$,
$\nu_{\mu}$,
$\nu_{\tau}$
are superpositions of three massive neutrinos
$\nu_1$,
$\nu_2$,
$\nu_3$
with respective masses
$m_1$,
$m_2$,
$m_3$
(see Ref.~\cite{Giunti:2007ry}).
There are two independent squared-mass differences,
the small solar
$\Delta{m}^2_{\text{SOL}} \simeq 7.5 \times 10^{-5} \, \text{eV}^2$
and the larger atmospheric
$\Delta{m}^2_{\text{ATM}} \simeq 2.3 \times 10^{-3} \, \text{eV}^2$,
which can be interpreted as
$
\Delta{m}^2_{\text{SOL}}
=
\Delta{m}^2_{21}
$
and
$
\Delta{m}^2_{\text{ATM}}
=
|\Delta{m}^2_{31}|
\simeq
|\Delta{m}^2_{32}|
$,
with
$\Delta{m}^2_{kj}=m_k^2-m_j^2$
(see Refs.\cite{Tortola:2012te,Fogli:2012ua,GonzalezGarcia:2012sz}).

The completeness of the $3\nu$ mixing paradigm has been challenged by
the following indications in favor of short-baseline neutrino oscillations,
which require the existence of at least one additional squared-mass difference,
$\Delta{m}^2_{\text{SBL}} \gg \Delta{m}^2_{\text{ATM}}$
(see the review in Ref.~\cite{Gariazzo:2015rra}):

\begin{enumerate}

\renewcommand{\labelenumi}{\theenumi.}
\renewcommand{\theenumi}{\arabic{enumi}}

\item
The reactor antineutrino anomaly
\cite{Mention:2011rk},
which is an about $2.8\sigma$ deficit of the rate of $\bar\nu_{e}$ observed in several
short-baseline reactor neutrino experiments
in comparison with that expected from the calculation of
the reactor neutrino fluxes
\cite{Mueller:2011nm,Huber:2011wv}.

\item
The Gallium neutrino anomaly
\cite{Abdurashitov:2005tb,Laveder:2007zz,Giunti:2006bj,Giunti:2010zu,Giunti:2012tn},
consisting in a short-baseline disappearance of $\nu_{e}$
measured in the
Gallium radioactive source experiments
GALLEX
\cite{Kaether:2010ag}
and
SAGE
\cite{Abdurashitov:2009tn}
with a statistical significance of about $2.9\sigma$.

\item
The
LSND experiment,
in which a signal of short-baseline
$\bar\nu_{\mu}\to\bar\nu_{e}$
oscillations has been observed
with a statistical significance of about $3.8\sigma$
\cite{Athanassopoulos:1995iw,Aguilar:2001ty}.

\end{enumerate}

The additional squared-mass difference
$\Delta{m}^2_{\text{SBL}}$
requires the existence of at least one massive neutrino
$\nu_4$
in addition to the three standard massive neutrinos
$\nu_1$,
$\nu_2$,
$\nu_3$.
Since from the LEP measurement of the invisible width of the $Z$ boson
we know that there are only three active neutrinos
(see Ref.~\cite{Giunti:2007ry}),
in the flavor basis the additional massive neutrinos correspond to
sterile neutrinos
\cite{Pontecorvo:1968fh},
which do not have standard weak interactions.

Sterile neutrinos are singlets of the Standard Model gauge symmetries
which can couple to the active neutrinos through the Lagrangian mass term.
In practice there are bounds on the active-sterile mixing,
but there is no bound on the number of sterile neutrinos and on their mass scales.
Therefore the existence of sterile neutrinos is investigated at different mass scales.
This review is devoted to the discussion of sterile neutrinos at the eV scale,
which can explain the indications in favor
of short-baseline neutrino oscillations listed above.
However, there are other very interesting possibilities which are
under study:
very light sterile neutrinos at a mass scale smaller than 0.1 eV, which could affect the oscillations of solar
\cite{deHolanda:2003tx,deHolanda:2010am,Das:2009kw}
and reactor
\cite{Kang:2013zma,Bakhti:2013ora,Palazzo:2013bsa,Girardi:2014gna,Girardi:2014wea,DiIura:2014csa,An:2014bik}
neutrinos;
sterile neutrinos at the keV scale,
which could constitute warm dark matter according to the Neutrino Minimal Standard Model ($\nu$MSM)
\cite{Asaka:2005an,Asaka:2005pn,Asaka:2006ek,Asaka:2006rw,Asaka:2006nq}
(see also the reviews in Refs.~\cite{Boyarsky:2009ix,Kusenko:2009up,Drewes:2013gca,Boyarsky:2012rt});
sterile neutrinos at the MeV scale
\cite{Kusenko:2004qc,Gelmini:2008fq,Masip:2012ke,Ishida:2014wqa};
sterile neutrinos at the electroweak scale
\cite{Antusch:2015mia,Deppisch:2015qwa}
or above it
\cite{Antusch:2014woa,Deppisch:2015qwa},
whose effects may be seen at LHC and other high-energy colliders.
Let us also note that there are several interesting models
with sterile neutrinos at different mass scales
\cite{Ghosh:2010hy,Barger:2010iv,Grossman:2010iq,deGouvea:2011zz,Chen:2011ai,Barry:2011fp,Fong:2011xh,Zhang:2011vh,Patra:2012ur,Nemevsek:2012iq,Dev:2012bd,Duerr:2013opa,Rodejohann:2014eka,Rosner:2014cha,Adhikari:2014nea,Merle:2014eja,deGouvea:2015pea}.

The possible existence of sterile neutrinos
is very interesting, because they are new particles which could
give us precious information on the physics beyond the Standard Model
(see Refs.~\cite{Volkas:2001zb,Mohapatra:2006gs}).
The existence of light sterile neutrinos is also very important for astrophysics
(see Ref.~\cite{Diaferio:2012zh})
and cosmology
(see Refs.~\cite{Lesgourgues-Mangano-Miele-Pastor-2013,RiemerSorensen:2013ih,Archidiacono:2013fha,Lesgourgues:2014zoa,Gariazzo:2015rra}).

In this review, we consider
3+1
\cite{Okada:1996kw,Bilenky:1996rw,Bilenky:1999ny,Maltoni:2004ei}
and
3+2
\cite{Sorel:2003hf,Karagiorgi:2006jf,Maltoni:2007zf,Karagiorgi:2009nb,Donini:2012tt},
neutrino mixing schemes
in which there are one or two additional massive neutrinos at the eV scale\footnote{
In the literature one can also find studies of the
3+3
\cite{Maltoni:2007zf,Rahman:2015hna},
3+1+1
\cite{Nelson:2010hz,Fan:2012ca,Kuflik:2012sw,Huang:2013zga,Giunti:2013aea},
and
1+3+1
\cite{Kopp:2011qd,Kopp:2013vaa}
schemes.
}
and
the masses of the three standard massive neutrinos are much smaller.
We do not consider schemes in which $\Delta m^2_{\text{SBL}}$
is obtained with one or more very light (or massless)
non-standard massive neutrinos and the three standard massive neutrinos have almost degenerate masses at the eV scale
(e.g., the 1+3, 1+3+1 and 2+3 schemes),
because this possibility is strongly disfavored by cosmological measurements
\cite{Ade:2015xua}
and by the experimental bound on
neutrinoless double-$\beta$ decay
(assuming that massive neutrinos are Majorana particles;
see Ref.~\cite{Bilenky:2014uka}).

In the 3+1 scheme,
the effective probability of
$\nua{\alpha}\to\nua{\beta}$
transitions in short-baseline experiments has the two-neutrino-like form
\cite{Bilenky:1996rw}
\begin{equation}
P_{\nua{\alpha}\to\nua{\beta}}
=
\delta_{\alpha\beta}
-
4 |U_{\alpha4}|^2 \left( \delta_{\alpha\beta} - |U_{\beta4}|^2 \right)
\sin^2\!\left( \dfrac{\Delta{m}^2_{41}L}{4E} \right)
,
\label{pab}
\end{equation}
where $U$ is the mixing matrix,
$L$ is the source-detector distance,
$E$ is the neutrino energy and
$\Delta{m}^2_{41} = m_{4}^2 - m_{1}^2 = \Delta{m}^2_{\text{SBL}} \sim 1 \, \text{eV}^2$.
The electron and muon neutrino and antineutrino appearance and disappearance
in short-baseline experiments
depend on
$|U_{e4}|^2$ and $|U_{\mu4}|^2$,
which
determine the amplitude
$\sin^22\vartheta_{e\mu} = 4 |U_{e4}|^2 |U_{\mu4}|^2$
of
$\nua{\mu}\to\nua{e}$
transitions,
the amplitude
$\sin^22\vartheta_{ee} = 4 |U_{e4}|^2 \left( 1 - |U_{e4}|^2 \right)$
of
$\nua{e}$
disappearance,
and
the amplitude
$\sin^22\vartheta_{\mu\mu} = 4 |U_{\mu4}|^2 \left( 1 - |U_{\mu4}|^2 \right)$
of
$\nua{\mu}$
disappearance.

Since the oscillation probabilities of neutrinos and antineutrinos are related by
a complex conjugation of the elements of the mixing matrix
(see Ref.~\cite{Giunti:2007ry}),
the effective probabilities of short-baseline
$\nu_{\mu}\to\nu_{e}$ and $\bar\nu_{\mu}\to\bar\nu_{e}$
transitions are equal.
Hence,
the 3+1 scheme cannot explain a possible CP-violating difference of
$\nu_{\mu}\to\nu_{e}$ and $\bar\nu_{\mu}\to\bar\nu_{e}$
transitions in short-baseline experiments.
In order to allow this possibility,
one must consider schemes with more than one sterile neutrino.
In the 3+2 scheme
there are four additional effective mixing parameters in short-baseline experiments:
$\Delta{m}^2_{51}\geq\Delta{m}^2_{41}$,
$|U_{e5}|^2$, $|U_{\mu5}|^2$
and
$\eta = \text{arg}\left[U_{e4}^*U_{\mu4}U_{e5}U_{\mu5}^*\right]$
(see Refs.~\cite{Conrad:2012qt,Gariazzo:2015rra}).
Since the complex phase $\eta$ appears with different signs in
the effective 3+2 probabilities of short-baseline
$\nu_{\mu}\to\nu_{e}$ and $\bar\nu_{\mu}\to\bar\nu_{e}$
transitions, it can generate measurable CP violations.

\section{Global fits of short-baseline data}
\label{sec:global}

Several analyses of
short-baseline neutrino oscillation data
have been done after the discovery of the LSND anomaly in the middle 90's
\cite{GomezCadenas:1995sj,Goswami:1995yq,Okada:1996kw,Bilenky:1996rw,Bilenky:1999ny,Giunti:2000wt,Barger:2000ch,Peres:2000ic,Grimus:2001mn,Strumia:2002fw,Maltoni:2002xd,Foot:2002tf,Giunti:2003cf,Sorel:2003hf,Barger:2003xm,Maltoni:2007zf,Goswami:2007kv,Schwetz:2007cd,Karagiorgi:2009nb,Akhmedov:2010vy,Nelson:2010hz}.
The interest in short-baseline neutrino oscillations
was renewed after the discovery in 2006 of the Gallium neutrino anomaly
\cite{Abdurashitov:2005tb,Laveder:2007zz,Giunti:2006bj,Acero:2007su,Giunti:2007xv,Giunti:2009zz,Giunti:2010wz,Giunti:2010zu,Giunti:2012tn,Giunti:2010zs,Giunti:2010jt,Giunti:2010uj}
and especially after the discovery in 2011 of the reactor antineutrino anomaly
\cite{Mention:2011rk,Kopp:2011qd,Donini:2011jh,Conrad:2011ce,Giunti:2011gz,Giunti:2011hn,Giunti:2011cp,Karagiorgi:2012kw,Kuflik:2012sw,Donini:2012tt,Archidiacono:2012ri,Giunti:2012tn,Giunti:2012bc,Conrad:2012qt,Kopp:2013vaa,Giunti:2013aea,Gariazzo:2015rra}.

\begin{table*}[t]
\begin{center}
\begin{tabular}{c|cccc|cc}
					&3+1									&3+1									&3+1									&3+1									&3+2									&3+2\\
					&TotGLO									&PrGLO								&noMB									&noLSND									&TotGLO									&PrGLO\\
\hline
$(\chi^{2}_{\text{min}})_{\text{GLO}}$	&306.0			&276.3			&251.2			&291.3			&299.6			&271.1			\\
$\text{NDF}_{\text{GLO}}$		&268			&262			&230			&264			&264			&258			\\
$\text{GoF}_{\text{GLO}}$		& 5\%		&26\%		&16\%		&12\%		& 7\%		&28\%		\\
\hline
$(\chi^{2}_{\text{min}})_{\text{APP}}$	&98.9			&77.0			&50.9			&91.8			&86.0			&69.6			\\
$(\chi^{2}_{\text{min}})_{\text{DIS}}$	&194.4			&194.4			&194.4			&194.4			&192.9			&192.9			\\
	$\Delta\chi^{2}_{\text{PG}}$	&13.0		&5.3		&6.2		&5.3		&20.7		&8.6		\\
	$\text{NDF}_{\text{PG}}$	&2		&2		&2		&2		&4		&4		\\
	$\text{GoF}_{\text{PG}}$	&0.1\%	& 7\%	& 5\%	& 7\%	&0.04\%	& 7\%	\\
\hline
$\Delta\chi^{2}_{\text{NO}}$		&$49.2$		&$47.7$		&$48.1$		&$11.4$		&$55.7$		&$52.9$		\\
	$\text{NDF}_{\text{NO}}$	&$3$		&$3$		&$3$		&$3$		&$7$		&$7$		\\
	$n\sigma_{\text{NO}}$		&$6.4\sigma$	&$6.3\sigma$	&$6.4\sigma$	&$2.6\sigma$	&$6.1\sigma$	&$5.9\sigma$	\\
\end{tabular}
\end{center}
\caption{ \label{tab:chi}
Results of the global fit of short-baseline data
taking into account all MiniBooNE data (TotGLO),
only the MiniBooNE data above 475 MeV (PrGLO),
without MiniBooNE data (noMB)
and without LSND data (noLSND)
in the
3+1 and 3+2 schemes.
The first three lines give
the minimum $\chi^{2}$ ($(\chi^{2}_{\text{min}})_{\text{GLO}}$),
the number of degrees of freedom ($\text{NDF}_{\text{GLO}}$) and
the goodness-of-fit ($\text{GoF}_{\text{GLO}}$)
of the global fit (GLO).
The following five lines give the quantities
relevant for the appearance-disappearance (APP-DIS) parameter goodness-of-fit (PG)
\protect\cite{Maltoni:2003cu}.
The last three lines give
the difference $\Delta\chi^{2}_{\text{NO}}$
between the $\chi^{2}$ without short-baseline oscillations
and $(\chi^{2}_{\text{min}})_{\text{GLO}}$,
the corresponding difference of number of degrees of freedom ($\text{NDF}_{\text{NO}}$)
and the resulting
number of $\sigma$'s ($n\sigma_{\text{NO}}$) for which the absence of oscillations is disfavored.
}
\end{table*}

Here we review the results
of the global fit of short-baseline neutrino oscillation
data presented in Ref.~\cite{Gariazzo:2015rra},
in which the data of the following three groups of experiments
have been considered:

\begin{enumerate}

\renewcommand{\labelenumi}{(\theenumi)}
\renewcommand{\theenumi}{\Alph{enumi}}

\item
The
$\nua{\mu}\to\nua{e}$
appearance data of the
LSND \cite{Aguilar:2001ty},
MiniBooNE \cite{Aguilar-Arevalo:2013pmq},
BNL-E776 \cite{Borodovsky:1992pn},
KARMEN \cite{Armbruster:2002mp},
NOMAD \cite{Astier:2003gs},
ICARUS \cite{Antonello:2013gut}
and
OPERA \cite{Agafonova:2013xsk}
experiments\footnote{
The correct but more complicated analysis of
the ICARUS and OPERA data
presented in Ref.~\cite{Palazzo:2015wea}
(see also Ref.~\cite{Donini:2007yf})
have not been considered
because it would not change significantly the results of the global fits.
}.

\item
The following
$\nua{e}$
disappearance data:
1)
the data of the
Bugey-4 \cite{Declais:1994ma},
ROVNO91 \cite{Kuvshinnikov:1990ry},
Bugey-3 \cite{Declais:1995su},
Gosgen \cite{Zacek:1986cu},
ILL \cite{Hoummada:1995zz},
Krasnoyarsk \cite{Vidyakin:1990iz},
Rovno88 \cite{Afonin:1988gx},
SRP \cite{Greenwood:1996pb},
Chooz \cite{Apollonio:2002gd},
Palo Verde \cite{Boehm:2001ik},
Double Chooz \cite{Abe:2014bwa}, and
Daya Bay \cite{An:2015nua}
reactor antineutrino experiments
with the new theoretical fluxes
\cite{Mueller:2011nm,Huber:2011wv,Mention:2011rk,Abazajian:2012ys};
2)
the data of the
GALLEX
\cite{Kaether:2010ag}
and
SAGE
\cite{Abdurashitov:2009tn}
Gallium radioactive source experiments
with the statistical method discussed in Ref.~\cite{Giunti:2010zu},
considering the recent
${}^{71}\text{Ga}({}^{3}\text{He},{}^{3}\text{H}){}^{71}\text{Ge}$
cross section measurement in Ref.~\cite{Frekers:2011zz};
3)
the solar neutrino constraint on $\sin^{2}2\vartheta_{ee}$
\cite{Giunti:2009xz,Palazzo:2011rj,Palazzo:2012yf,Giunti:2012tn,Palazzo:2013me};
4)
the
KARMEN \cite{Armbruster:1998uk}
and
LSND \cite{Auerbach:2001hz}
$\nu_{e} + {}^{12}\text{C} \to {}^{12}\text{N}_{\text{g.s.}} + e^{-}$
scattering data \cite{Conrad:2011ce},
with the method discussed in Ref.~\cite{Giunti:2011cp}.

\item
The constraints on
$\nua{\mu}$
disappearance obtained from
the data of the
CDHSW experiment \cite{Dydak:1983zq},
from the analysis \cite{Maltoni:2007zf} of
the data of
atmospheric neutrino oscillation experiments\footnote{
The analysis of the IceCube data
\cite{Barger:2011rc,Razzaque:2012tp,Esmaili:2012nz,Esmaili:2013vza,Lindner:2015iaa},
which could give a marginal contribution,
have not been considered
because it is too complicated and subject to large uncertainties.
},
from the analysis \cite{Hernandez:2011rs,Giunti:2011hn} of the
MINOS neutral-current data \cite{Adamson:2011ku}
and from the analysis of the
SciBooNE-MiniBooNE data
neutrino \cite{Mahn:2011ea} and antineutrino \cite{Cheng:2012yy} data.

\end{enumerate}

The MiniBooNE data require a special treatment,
because they show an anomalous excess in the low-energy bins
\cite{AguilarArevalo:2008rc,Aguilar-Arevalo:2013pmq}
which, as explained later,
induces a tension in the global analysis of the data of
short-baseline neutrino oscillation experiments
\cite{Giunti:2011hn,Giunti:2011cp}.
Hence, we will discuss two types of global fits:
``total'' (TotGLO) and ``pragmatic'' (PrGLO).
In the total fits all the data listed above of short-baseline neutrino oscillation experiments
are taken into account.
In the pragmatic fits \cite{Giunti:2013aea}
the anomalous low-energy bins of the MiniBooNE experiment
\cite{AguilarArevalo:2008rc,Aguilar-Arevalo:2013pmq}
are omitted.

\begin{figure*}[t]
\null
\hfill
\includegraphics*[width=0.49\linewidth]{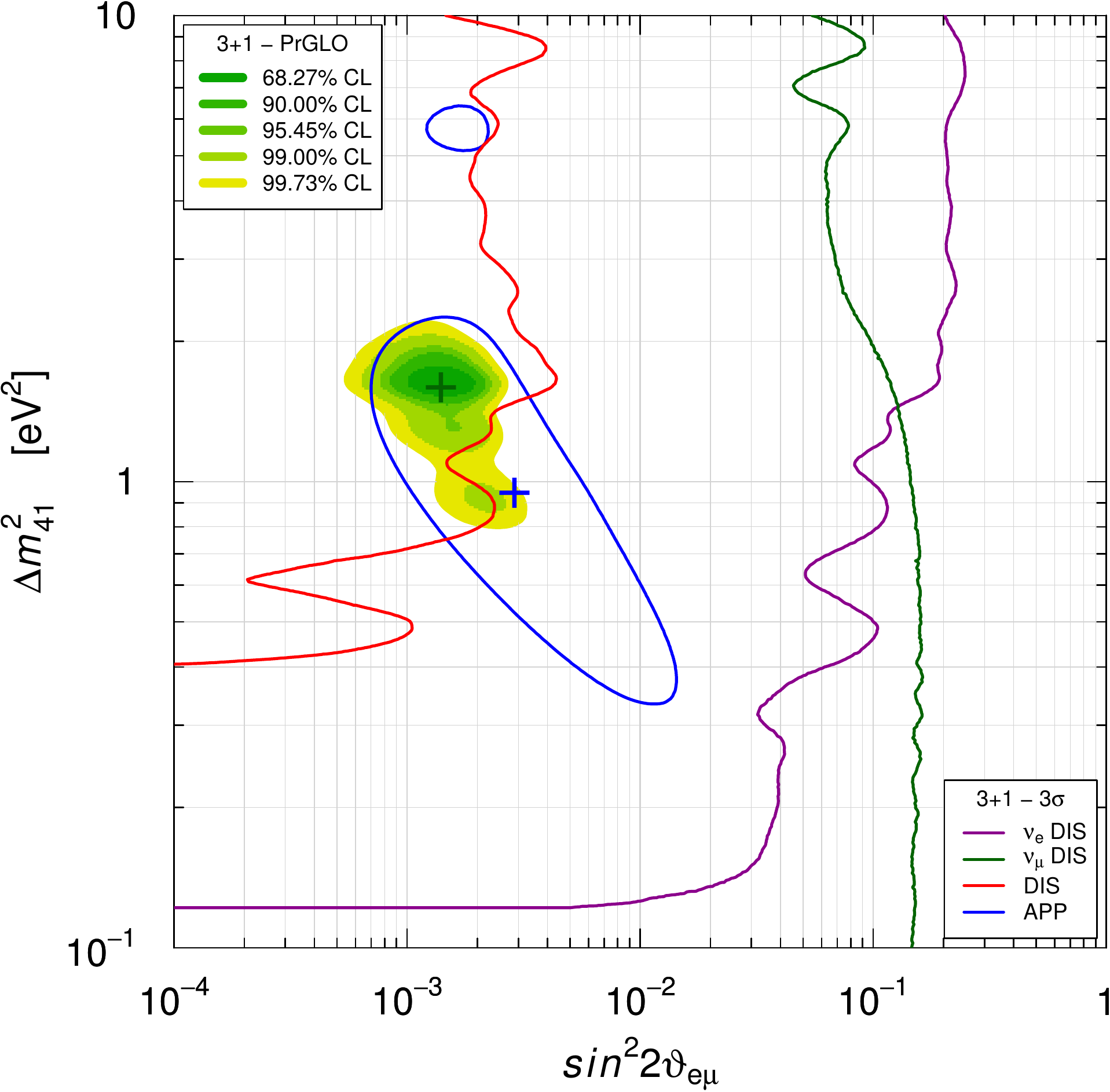}
\hfill
\includegraphics*[width=0.49\linewidth]{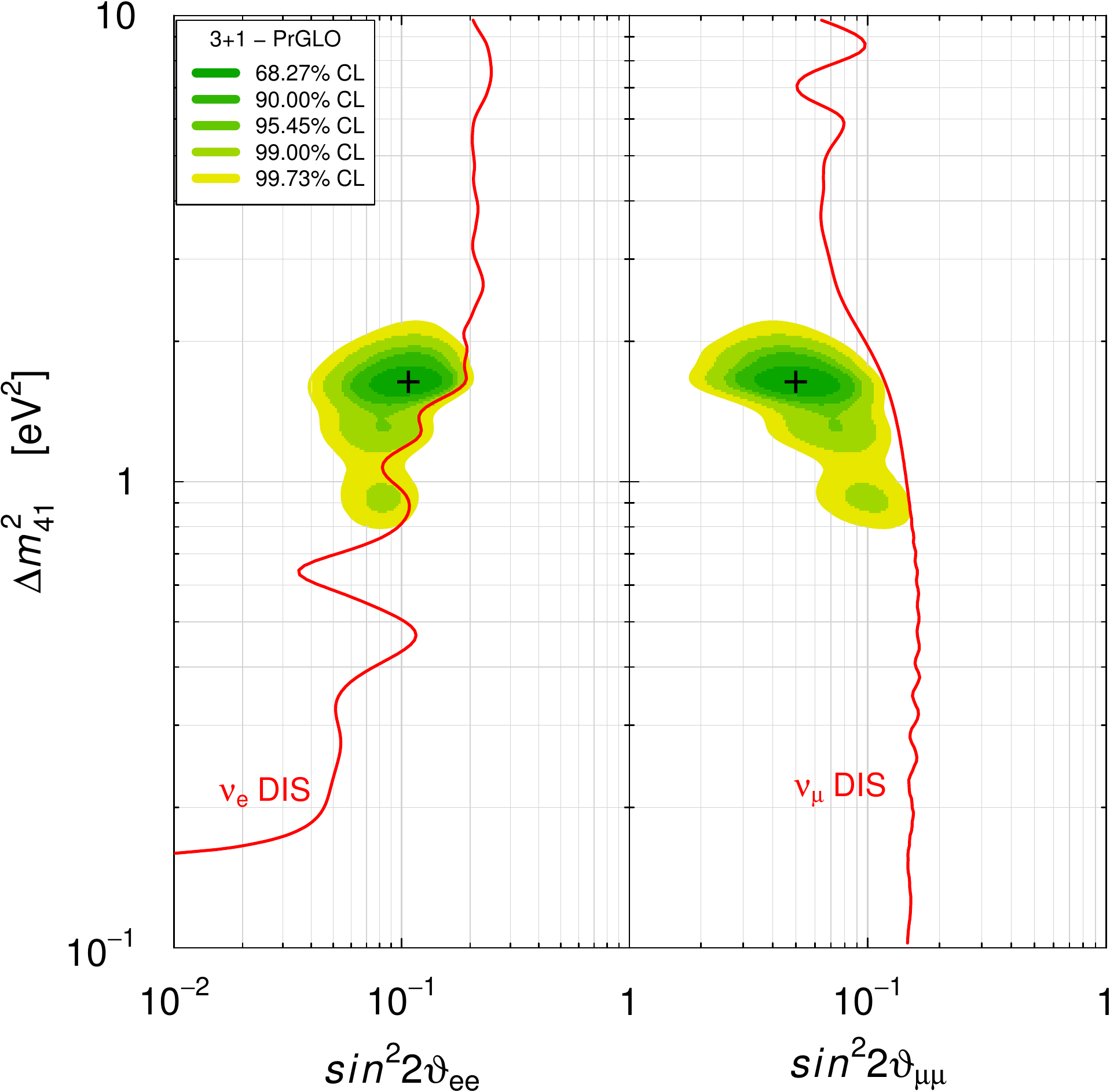}
\hfill
\null
\caption{ \label{fig:glo}
Allowed regions in the
$\sin^{2}2\vartheta_{e\mu}$--$\Delta{m}^{2}_{41}$,
$\sin^{2}2\vartheta_{ee}$--$\Delta{m}^{2}_{41}$
and
$\sin^{2}2\vartheta_{\mu\mu}$--$\Delta{m}^{2}_{41}$
planes
obtained in the pragmatic 3+1 global fit PrGLO
of short-baseline neutrino oscillation data
compared with the $3\sigma$ allowed regions
obtained from
$\protect\nua{\mu}\to\protect\nua{e}$
short-baseline appearance data (APP)
and the $3\sigma$ constraints obtained from
$\protect\nua{e}$
short-baseline disappearance data ($\nu_{e}$ DIS),
$\protect\nua{\mu}$
short-baseline disappearance data ($\nu_{\mu}$ DIS)
and the
combined short-baseline disappearance data (DIS).
The best-fit points of the global (PrGLO) and APP fits are indicated by crosses.
}
\end{figure*}

Table~\ref{tab:chi}
summarizes the statistical results obtained
from global fits of the data above
in the
3+1 and 3+2 schemes.
Besides the total and pragmatic fits there is also a 3+1-noMB fit without MiniBooNE data
and
a 3+1-noLSND fit without LSND data
which are explained below.

From Tab.~\ref{tab:chi},
one can see that in all fits which include the LSND data
the absence of short-baseline oscillations
is nominally disfavored by about $6\sigma$,
because the improvement of the $\chi^2$ with short-baseline oscillations
is much larger than the number of oscillation parameters.

In both the
3+1 and 3+2 schemes
the goodness-of-fit in the total analysis is significantly worse than that in the pragmatic analysis
and the appearance-disappearance parameter goodness-of-fit is much worse.
This result confirms the fact that the MiniBooNE low-energy anomaly
is incompatible with neutrino oscillations,
because it would require a small value of $\Delta{m}^2_{41}$
and a large value of $\sin^22\vartheta_{e\mu}$
\cite{Giunti:2011hn,Giunti:2011cp},
which are excluded by the data of other experiments
(see Ref.~\cite{Giunti:2013aea} for further details)\footnote{
One could fit the three anomalous MiniBooNE low-energy bins
in a 3+2 scheme \cite{Conrad:2012qt}
by considering the appearance data without the
ICARUS \cite{Antonello:2013gut}
and
OPERA \cite{Agafonova:2013xsk}
constraints,
but the required large transition probability is excluded
by the disappearance data.
}.
Note that the appearance-disappearance tension
in the 3+2-TotGLO fit is even worse than that in the 3+1-TotGLO fit,
since the $\Delta\chi^{2}_{\text{PG}}$ is so much larger that it cannot be compensated
by the additional degrees of freedom\footnote{
This behavior has been explained in Ref.~\cite{Archidiacono:2013xxa}.
It was found also in the analysis presented in Ref.~\cite{Kopp:2013vaa}.
}.
Therefore,
we think that it is very likely that the MiniBooNE low-energy anomaly
has an explanation which is different from neutrino oscillations\footnote{
There is however the possibility that at least some part of the MiniBooNE low-energy anomaly
may be explained by taking into account nuclear effects in the energy reconstruction
\cite{Martini:2012fa,Martini:2012uc}.
}.
The cause of the MiniBooNE low-energy excess of $\nu_{e}$-like events
is going to be investigated in the MicroBooNE experiment at Fermilab
\cite{Szelc:2015dga},
which is a large Liquid Argon Time Projection Chamber (LArTPC)
in which electrons and photons can be distinguished\footnote{
In the MiniBooNE  mineral-oil Cherenkov detector
$\nu_{e}$-induced events
cannot be distinguished from
$\nu_{\mu}$-induced events which produce only a visible photon
(for example neutral-current $\pi^{0}$ production in which only one of the two decay photons is visible).
}.

In the following we adopt the ``pragmatic approach'' advocated in Ref.~\cite{Giunti:2013aea}
which considers the PrGLO fits,
without the anomalous MiniBooNE low-energy bins,
as more reliable than the TotGLO fits,
which include the anomalous MiniBooNE low-energy bins.

The 3+2 mixing scheme
was considered to be interesting in 2010
when the MiniBooNE neutrino
\cite{AguilarArevalo:2008rc}
and antineutrino
\cite{AguilarArevalo:2010wv}
data showed a CP-violating tension,
but
this tension almost disappeared in the final MiniBooNE data
\cite{Aguilar-Arevalo:2013pmq}.
In fact, from Tab.~\ref{tab:chi}
one can see that there is little improvement of the 3+2-PrGLO fit
with respect to the 3+1-PrGLO fit,
in spite of the four additional parameters and the additional possibility of CP violation.
Moreover,
the p-value obtained by restricting the 3+2 scheme to 3+1
disfavors the 3+1 scheme only at
$1.1\sigma$.
Therefore,
we think that considering the larger complexity of the 3+2 scheme
is not justified by the data
and in the following we consider only the
3+1 mixing scheme.

Figure~\ref{fig:glo}
shows the allowed regions in the
$\sin^{2}2\vartheta_{e\mu}$--$\Delta{m}^{2}_{41}$,
$\sin^{2}2\vartheta_{ee}$--$\Delta{m}^{2}_{41}$ and
$\sin^{2}2\vartheta_{\mu\mu}$--$\Delta{m}^{2}_{41}$
planes
obtained in the 3+1-PrGLO fit.
These regions are relevant, respectively, for
$\nua{\mu}\to\nua{e}$ appearance,
$\nua{e}$ disappearance and
$\nua{\mu}$ disappearance
searches.
Figure~\ref{fig:glo}
shows also the region allowed by $\nua{\mu}\to\nua{e}$ appearance data
and
the constraints from
$\nua{e}$ disappearance and
$\nua{\mu}$ disappearance data.
One can see that the combined disappearance constraint
in the $\sin^{2}2\vartheta_{e\mu}$--$\Delta{m}^{2}_{41}$ plane
excludes a large part of the region allowed by $\nua{\mu}\to\nua{e}$ appearance data,
leading to the well-known
appearance-disappearance tension
\cite{Kopp:2011qd,Giunti:2011gz,Giunti:2011hn,Giunti:2011cp,Conrad:2012qt,Archidiacono:2012ri,Archidiacono:2013xxa,Kopp:2013vaa,Giunti:2013aea,Gariazzo:2015rra,Giunti:2015mwa}
quantified by the parameter goodness-of-fit in Tab.~\ref{tab:chi}.
The best-fit values of the oscillation parameters are
$(\Delta{m}^{2}_{41})_{\text{bf}} = 1.6 \, \text{eV}^2$,
$(|U_{e4}|^2)_{\text{bf}} = 0.028$,
$(|U_{\mu4}|^2)_{\text{bf}} = 0.013$,
which imply
$(\sin^{2}2\vartheta_{e\mu})_{\text{bf}} = 0.0014$,
$(\sin^{2}2\vartheta_{ee})_{\text{bf}} = 0.11$ and
$(\sin^{2}2\vartheta_{\mu\mu})_{\text{bf}} = 0.050$.

It is interesting to investigate what are the
impacts of the MiniBooNE and LSND experiments
on the global analysis of short-baseline neutrino oscillation data.
With this aim,
we consider two additional 3+1 fits:
a 3+1-noMB fit without MiniBooNE data
and
a 3+1-noLSND fit without LSND data.
From Tab.~\ref{tab:chi}
one can see that the results of the
3+1-noMB fit are similar to those of the
3+1-PrGLO fit
and the nominal exclusion of the case of no-oscillations remains at the level of $6\sigma$.
On the other hand,
in the 3+1-noLSND fit,
without LSND data,
the nominal exclusion of the case of no-oscillations drops dramatically to
$2.6\sigma$.
In fact,
in this case
the main indication in favor of short-baseline oscillations
is given by the reactor
and
Gallium
anomalies
which have a similar statistical significance
\cite{Giunti:2012tn}.
Therefore,
it is clear that the LSND experiment is still crucial for the indication in favor of short-baseline
$\bar\nu_{\mu}\to\bar\nu_{e}$
transitions.

\begin{table*}[t]
\begin{center}
\begin{tabular}{lccccc}
\hline
Project							& neutrino		& source		& $E$			& $L$		& status	\\
							& 			& 			& (MeV)			& (m)		&		\\
\hline
SAGE		\cite{Gavrin:2015aca}			& $\nu_e$		& $^{51}$Cr		& $0.75$		& $\lesssim1$	& in preparation\\
CeSOX		\cite{Borexino:2013xxa,Gaffiot:2015fva}	& $\bar\nu_e$		& $^{144}$Ce		& $1.8-3$		& $5-12$	& in preparation\\
CrSOX		\cite{Borexino:2013xxa}			& $\nu_e$		& $^{51}$Cr		& $0.75$		& $5-12$	& proposal	\\
Daya Bay	\cite{Dwyer:2011xs,Gao:2013tha}		& $\bar\nu_e$		& $^{144}$Ce		& $1.8-3$		& $1.5-8$	& proposal	\\
JUNO		\cite{An:2015jdp}			& $\bar\nu_e$		& $^{144}$Ce		& $1.8-3$		& $\lesssim32$	& proposal	\\
LENS		\cite{Agarwalla:2010gd}			& $\nu_{e},\bar\nu_e$	& $^{51}$Cr, $^6$He	& $0.75$, $\lesssim3.5$	& $\lesssim3$	& abandoned	\\
CeLAND		\cite{Gando:2013zoa}			& $\bar\nu_e$		& $^{144}$Ce		& $1.8-3$		& $\lesssim6$	& abandoned	\\
LENA		\cite{Novikov:2011gp}			& $\nu_e$		& $^{51}$Cr, $^{37}$Ar	& $0.75$, $0.81$	& $\lesssim90$	& abandoned	\\
\hline
\end{tabular}
\caption{Main features of new source experiments and their status according to our knowledge.}
\label{tab:source}
\end{center}
\end{table*}

\begin{table*}[t]
\begin{center}
\begin{tabular}{lccccc}
\hline
Project							& $P_{th}$	& $M_{target}$	& $L$		& Depth		& status	\\
							& (MW)		& (tons)	& (m)		& (m.w.e.)	&		\\
\hline
Nucifer (FRA)		\cite{Boireau:2015dda}		& $70$		& $0.8$		& $7$		& $13$		& operating	\\
Stereo (FRA)		\cite{Pequignot:2015rta}	& $57$		& $1.75$	& $9-12$	& $18$		& in preparation\\
DANSS (RUS)		\cite{Danilov:2014vra}		& $3000$	& $0.9$		& $10-12$	& $50$		& in preparation\\
SoLid (BEL)		\cite{Ryder:2015sma}		& $45-80$	& $3$		& $6-8$		& $10$		& in preparation\\
PROSPECT (USA)		\cite{Ashenfelter:2015uxt}	& $85$		& $3,10$	& $7-12,15-19$	& few		& in preparation\\
NEOS (KOR)		\cite{Kim:2015qlu}		& $16400$	& $1$		& $25$		& $10-23$	& in preparation\\
Neutrino-4 (RUS)	\cite{Serebrov:2013yaa}		& $100$		& $1.5$		& $6-11$	& $10$		& proposal	\\
Poseidon (RUS)		\cite{Derbin:2012kf}		& $100$		& $3$		& $5-8$		& $15$		& proposal	\\
Hanaro (KOR)		\cite{Yeo:2014spa}		& $30$		& $0.5$		& $6$		& few		& proposal	\\
CARR (CHN)		\cite{Guo:2013sea}		& $60$		& $\sim1$	& $7,11$	& few		& proposal	\\
\hline
\end{tabular}
\caption{Main features of new reactor experiments and their status according to our knowledge.}
\label{tab:reactor}
\end{center}
\end{table*}

\section{Experimental perspectives}
\label{sec:perspectives}

There is an impressive program of many experimental projects
which will explore the existence of light sterile neutrinos at the eV scale
in the next years
(see also the reviews in
Refs.~\cite{Lhuillier:2014mna,Katori:2014vka,Lasserre:2014ita,Caccianiga:2015ega,Lhuillier:2015fga,Spitz:2015gga,Lasserre:2015eva}).
It is convenient to divide them in the following categories.

\subsection{$\protect\nua{e}$ disappearance experiments}
\label{sub:nuedis}

The aim of these experiments is to reveal short-baseline oscillations in a robust way
by measuring distortions of the neutrino spectrum
or variations of the flavor neutrino detection probability as a function of distance.
They can be divided in the following subcategories.

\begin{description}

\item[Source experiments.]
These experiments use radioactive sources
of $\nu_{e}$ or $\bar\nu_{e}$
placed near or inside a large detector
\cite{Cribier:2011fv}.
Table~\ref{tab:source}
presents a list of the projects which have been proposed
(see also Ref.~\cite{Caccianiga:2015ega}).

In source experiments with monochromatic $\nu_{e}$'s generated by nuclear electron capture
(for example SAGE \cite{Gavrin:2015aca} and CrSOX \cite{Borexino:2013xxa}),
$\nu_{e}$ disappearance can be measured as a function of distance.
In source experiments with a continuous $\bar\nu_{e}$ spectrum generated by nuclear $\beta$ decay
(for example CeSOX \cite{Borexino:2013xxa,Gaffiot:2015fva})
also the distortions of the neutrino spectrum
can be measured.

\item[Reactor experiments.]
These experiments use a reactor $\bar\nu_{e}$ source with a detector placed at a distance of the order of 10 m.
There are several experiments in preparation,
as shown by the list in Tab.~\ref{tab:reactor}
(see also Ref.~\cite{Lhuillier:2015fga}).
They are planned to have a sufficient energy resolution in order to be sensitive to the distortions in the neutrino spectrum
due to the oscillations.
Some experiments
(for example Stereo \cite{Pequignot:2015rta})
will have a length which may allow to
observe the variations of the $\bar\nu_{e}$ survival probability as a function of distance.
Others use will use two detectors at different distances
(for example PROSPECT \cite{Ashenfelter:2013oaa} and CARR \cite{Guo:2013sea})
or a movable detector
(for example DANSS \cite{Danilov:2014vra}).

\item[Accelerator experiments.]
There have been proposals to use a future $\beta$-beam
\cite{Espinoza:2013dsa,Agarwalla:2009em,Bungau:2012ys}
or a low-energy neutrino factory
\cite{Giunti:2009en,Meloni:2010zr,Winter:2012sk,Adey:2014rfv}
to search for short-baseline
$\nu_{e}$ disappearance.

\end{description}

\begin{figure}[t]
\includegraphics*[width=\linewidth]{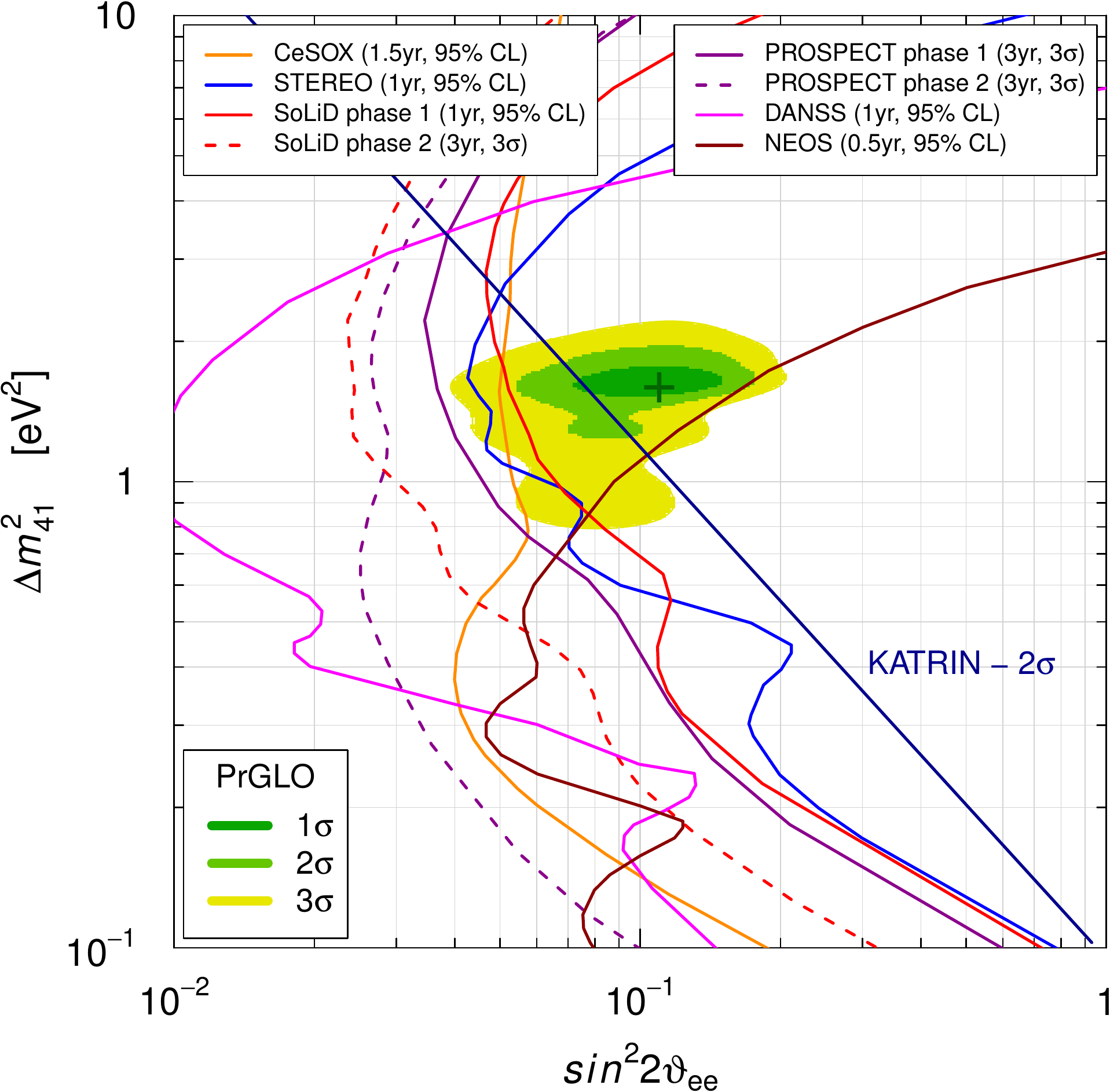}
\caption{ \label{fig:fut-see}
Comparison of the allowed region in the $\sin^{2}2\vartheta_{ee}$--$\Delta{m}^{2}_{41}$ plane
obtained in the pragmatic 3+1 global fit PrGLO
of short-baseline neutrino oscillation data
with the sensitivities of the
CeSOX \cite{Borexino:2013xxa,Gaffiot:2015fva} source experiment,
of the
Stereo \cite{Pequignot:2015rta},
SoLid \cite{Ryder:2015sma},
DANSS \cite{Danilov:2014vra} and
NEOS \cite{Kim:2015qlu}
reactor experiments
and of the KATRIN \cite{Mertens-TAUP2015} $\beta$-decay experiment.
}
\end{figure}

Figure~\ref{fig:fut-see}
shows the sensitivities
in the $\sin^{2}2\vartheta_{ee}$--$\Delta{m}^{2}_{41}$ plane
of the
CeSOX \cite{Borexino:2013xxa,Gaffiot:2015fva} source experiment
and
of the
Stereo \cite{Pequignot:2015rta},
SoLid \cite{Ryder:2015sma},
DANSS \cite{Danilov:2014vra} and
NEOS \cite{Kim:2015qlu}
reactor experiments
in comparison with the region allowed by the
pragmatic 3+1 global fit PrGLO.
One can see that these experiments should be able to check unambiguously
the indications of
short-baseline neutrino oscillations.

\begin{table*}[t]
\begin{center}
\begin{tabular}{lccccc}
\hline
Project						& $P$		& $M_{target}$	& $E$		& L		& status	\\
						& (MW)		& (tons)	& (MeV)		& (m)		&		\\
\hline
SBN (USA)		\cite{Antonello:2015lea}& $> 0.09$	& $112,89,476$	& $\sim 800$	& $110,470,600$	& in preparation\\
J-PARC MLF (JPN)	\cite{Ajimura:2015yux}	& $\sim 1$	& $50$		& $\sim 40$	& $20$		& proposal	\\
KPipe (JPN)		\cite{Axani:2015zxa}	& $\sim 1$	& $684$		& $\sim 236$	& $32-152$	& proposal	\\
nuPRISM (JPN)		\cite{Bhadra:2014oma}	& $\sim 1$	& $4000-8000$	& $200-1000$	& $1000-2000$	& proposal	\\
IsoDAR-KamLAND (JPN)	\cite{Abs:2015tbh}	& $0.6$		& $1000$	& $\sim 6.5$	& $10-22$	& proposal	\\
IsoDAR-JUNO (CHN)	\cite{An:2015jdp}	& $0.6$		& $20000$	& $\sim 6.5$	& $20-100$	& proposal	\\
OscSNS (USA)		\cite{Elnimr:2013wfa}	& $1.4$		& $450$		& $\sim 40$	& $50-70$	& proposal	\\
\hline
\end{tabular}
\caption{Main features of new accelerator experiments and their status according to our knowledge.}
\label{tab:accelerator}
\end{center}
\end{table*}

\begin{figure}[t]
\includegraphics*[width=\linewidth]{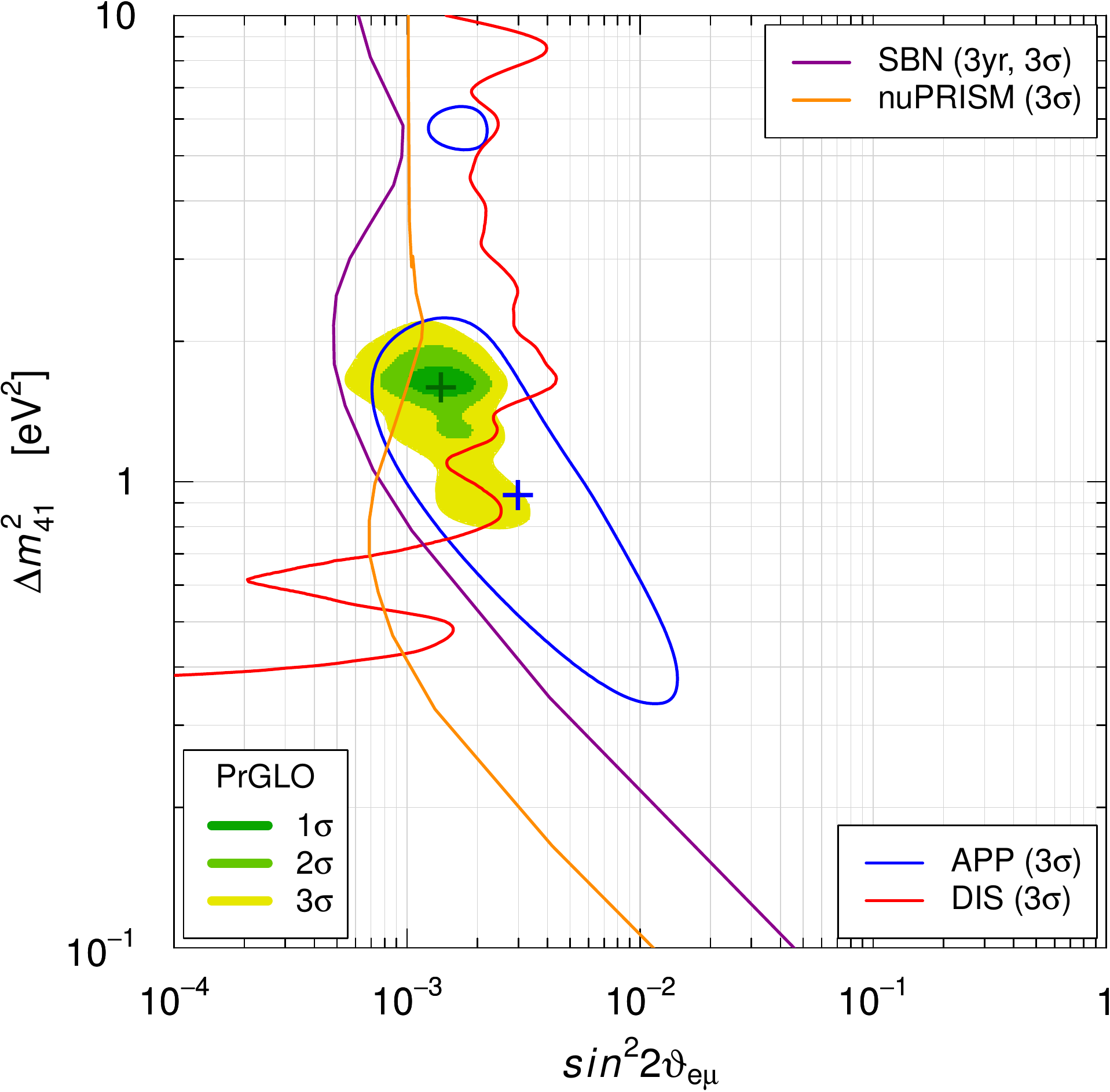}
\caption{ \label{fig:fut-sem}
Comparison of the allowed region in the $\sin^{2}2\vartheta_{e\mu}$--$\Delta{m}^{2}_{41}$ plane
obtained in the pragmatic 3+1 global fit PrGLO
of short-baseline neutrino oscillation data
with the sensitivities of the
SBN \cite{Antonello:2015lea}
and
nuPRISM \cite{Bhadra:2014oma}
accelerator experiments.
}
\end{figure}

\begin{figure}[t]
\includegraphics*[width=\linewidth]{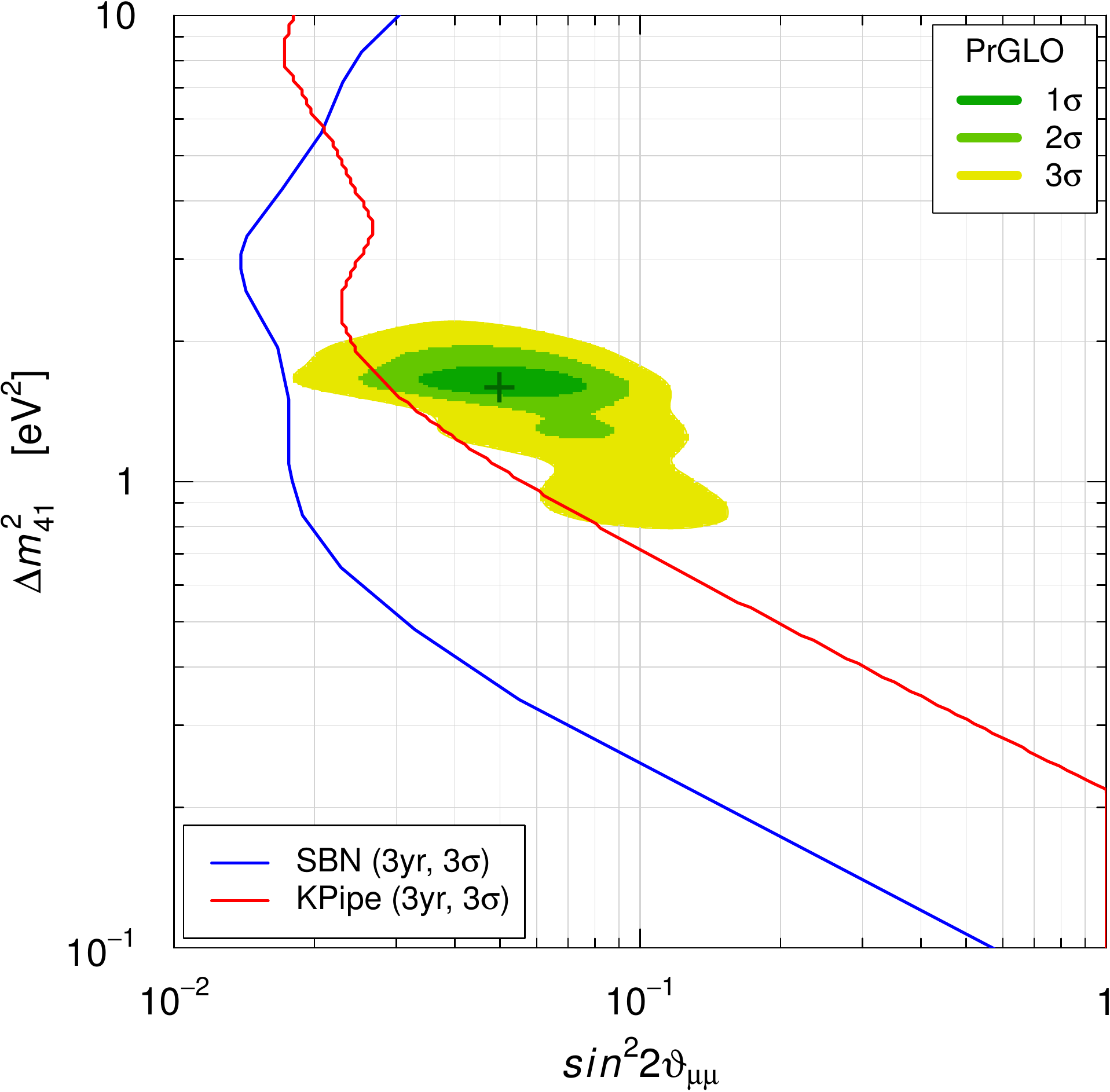}
\caption{ \label{fig:fut-smm}
Comparison of the allowed region in the $\sin^{2}2\vartheta_{\mu\mu}$--$\Delta{m}^{2}_{41}$ plane
obtained in the pragmatic 3+1 global fit PrGLO
of short-baseline neutrino oscillation data
with the sensitivities of the
SBN \cite{Antonello:2015lea}
and
KPipe \cite{Axani:2015zxa}
accelerator experiments.
}
\end{figure}

\subsection{$\protect\nua{\mu}\to\protect\nua{e}$ appearance experiments}
\label{sub:app}

The $\nua{\mu}\to\nua{e}$ appearance channel
can be explored in accelerator experiments with a beam of $\nua{\mu}$.
The main projects are listed in Tab.~\ref{tab:accelerator}
(see also Ref.~\cite{Spitz:2015gga}).
They aim at checking the short-baseline
$\bar\nu_{\mu}\to\bar\nu_{e}$
LSND signal
in both neutrino
($\nu_{\mu}\to\nu_{e}$)
and antineutrino
($\bar\nu_{\mu}\to\bar\nu_{e}$)
mode
(see also the low-energy neutrino factory studies in Refs.~\cite{Meloni:2010zr,Tunnell:2011ya,Adey:2014rfv}).

For accelerator experiments
a crucial ingredient for reaching a robust result is the presence of ``near'' and ``far'' detectors
(as, for example, in the SBN \cite{Antonello:2015lea} experiment,
where there will be even the ``middle'' MicroBooNE detector, albeit smaller than the near detector).
The near detector provides a normalization of the neutrino flux and cross section
which allows to measure the oscillations between the two detectors with small systematic uncertainty.

Figure~\ref{fig:fut-sem}
shows the sensitivities
in the $\sin^{2}2\vartheta_{e\mu}$--$\Delta{m}^{2}_{41}$ plane
of the
SBN \cite{Antonello:2015lea}
and
nuPRISM \cite{Bhadra:2014oma}
accelerator experiments,
which are expected to observe with a convincing statistical significance
$\nua{\mu}\to\nua{e}$
consistent with the LSND signal
if sterile neutrinos at the eV scale exist.

\subsection{$\protect\nua{\mu}$ disappearance experiments}
\label{sub:numudis}

The accelerator experiments in Tab.~\ref{tab:accelerator}
(see also the NESSiE proposal in Ref.~\cite{Stanco:2013dha}
and
the low-energy neutrino factory studies in Refs.~\cite{Meloni:2010zr,Tunnell:2011ya,Adey:2014rfv})
can measure also the short-baseline
$\nu_{\mu}$ and $\bar\nu_{\mu}$
disappearance which is necessarily associated with
$\nua{\mu}\to\nua{e}$
oscillations.
Let us emphasize that it is important to measure short-baseline
$\nu_{\mu}$ and $\bar\nu_{\mu}$
disappearance
for which so far there are only upper limits,
whereas the short-baseline
$\nu_{e}$ and $\bar\nu_{e}$
disappearance associated with
$\nua{\mu}\to\nua{e}$
oscillations
is given by the Gallium and reactor anomalies.
The consistency of the short-baseline neutrino oscillation scenario
with any number of sterile neutrinos requires
that also $\nu_{\mu}$ and $\bar\nu_{\mu}$
disappearance must be observed
\cite{Giunti:2015mwa}.

Figure~\ref{fig:fut-smm}
shows the sensitivities
in the $\sin^{2}2\vartheta_{\mu\mu}$--$\Delta{m}^{2}_{41}$ plane
of the
SBN \cite{Antonello:2015lea}
and
KPipe \cite{Axani:2015zxa}
accelerator experiments.
One can see that also $\nua{\mu}$ disappearance
should be observed if the short-baseline neutrino oscillations
indicated by the LSND, reactor and Gallium anomalies
really exist.

\subsection{Neutral-current measurements}
\label{sub:neutral}

In principle, measuring the neutral-current scattering of active neutrinos
is the best way to probe their disappearance into sterile states.
However, neutral-current measurements are extremely difficult,
because the only observable signal is the recoil of the target particle.

The signal can be enhanced at low neutrino energies by the coherent scattering on nuclei
\cite{Freedman:1973yd,Drukier:1983gj}
for which the cross section is approximately proportional to the square of the number of neutrons in the nucleus
(the proton contribution is suppressed by
$1 - 4 \sin^2\vartheta_{\text{W}} \ll 1$,
where $\vartheta_{\text{W}}$ is the weak mixing angle).
This process has not been observed so far, but it is actively searched for
\cite{Scholberg:2005qs,Barranco:2005yy,Anderson:2011bi,Vergados:2009ei,Akimov:2015nza}.
In the future it may lead to the direct measurement of
active-sterile transitions
\cite{Formaggio:2011jt,Anderson:2012pn,Dutta:2015nlo}.

\subsection{$\beta$-decay mass measurements}
\label{sub:bd}

The most sensitive experiments on the search of the effects of neutrino masses in $\beta$ decay
use the Tritium decay process\footnote{
Other methods are described in the reviews in Refs.~\cite{Drexlin:2013lha,Formaggio:2014ppa,Gastaldo:2014hga,Dragoun:2015oja,Nucciotti:2015rsl}
}
\begin{equation}
{}^{3}\text{H} \to
{}^{3}\text{He} + e^{-} + \bar\nu_{e}
.
\label{tritium}
\end{equation}
Non-zero neutrino masses distort the measurable spectrum of the emitted electron.
It is convenient to consider the Kurie function (see Ref.~\cite{Giunti:2007ry})
\begin{align}
K^2(T)
=
\left( Q - T_{e} \right)
\sum_{k}
\null & \null
|U_{ek}|^2
\sqrt{ \left( Q - T_{e} \right)^{2} - m_{k}^{2} }
\nonumber
\\
\null & \null
\times
\Theta(Q-T_{e}-m_{k})
,
\label{kurie1}
\end{align}
where $T_{e}$ is the electron kinetic energy,
$
Q
=
M_{{}^{3}\text{H}}
-
M_{{}^{3}\text{He}}
-
m_{e}
\simeq
18.574 \, \text{keV}
$
is the $Q$-value of the process,
and
$\Theta$ is the Heaviside step function.
Considering an experiment in which the energy resolution is such that
$m_{k} \ll Q-T_{e}$
for the three standard light neutrino masses ($k=1,2,3$),
the Kurie function can be approximated by
\begin{align}
K^2(T)
\simeq
\null & \null
\left( Q - T_{e} \right)
\sqrt{ \left( Q - T_{e} \right)^{2} - m_{\beta}^{2} }
\,
\Theta(Q-T_{e}-m_{\beta})
\nonumber
\\
\null & \null
+
\left( Q - T_{e} \right)
\sum_{k\geq4}
|U_{ek}|^2
\sqrt{ \left( Q - T_{e} \right)^{2} - m_{k}^{2} }
\nonumber
\\
\null & \null
\phantom{+
\left( Q - T_{e} \right)
\sum_{k\geq4}}
\times
\Theta(Q-T_{e}-m_{k})
,
\label{kurie2}
\end{align}
with the effective light neutrino mass $m_{\beta}$ given by
\begin{equation}
m_{\beta}^2 = \sum_{k=1}^{3} |U_{ek}|^2 m_{k}^2
.
\label{efnmass}
\end{equation}
Hence,
$m_{\beta}$ causes a distortion of the end-point of the electron kinetic energy spectrum
and
a heavy nonstandard neutrino mass $m_{k}$ with $k\geq4$ can be measured by observing
a kink of the kinetic energy spectrum of the emitted electron at
$Q-m_{k}$
below the end point
\cite{Shrock:1980vy,Schreckenbach:1983cg,Ohshima:1993pp,Mortara:1993iv,Farzan:2001cj,Farzan:2002zq,deGouvea:2006gz,Riis:2010zm,Giunti:2011cp}.
Recently,
the Mainz
\cite{Kraus:2012he}
and
Troitsk
\cite{Belesev:2012hx,Belesev:2013cba}
collaborations obtained upper bounds for the mixing factor $|U_{e4}|^2$
for $m_{4}^2 \gtrsim 10 \, \text{eV}^2$.
In the 3+1 scheme these bounds
imply an exclusion curve in the
$\sin^{2}2\vartheta_{ee}$--$\Delta{m}^{2}_{41}$ plane
for
$\Delta{m}^2_{41} \gtrsim 10 \, \text{eV}^2$
\cite{Giunti:2012bc},
which is well above the allowed region
obtained in the pragmatic 3+1 global fit PrGLO
shown in Fig.~\ref{fig:glo}.

The experiment KATRIN \cite{Mertens:2015ila},
which is under construction and is scheduled to start data taking in 2016,
will aim to reach a sensitivity of $0.2 \, \text{eV}$ at 90\% C.L. for $m_{\beta}$ in five years of running.
Some studies have been performed to analyze the sensitivity of the KATRIN experiment
to the effects of
heavy sterile neutrinos with keV-scale masses
\cite{Abdurashitov:2015jta,Mertens:2014nha,Barry:2014ika,Mertens-TAUP2015}
and
light eV-scale sterile neutrinos
\cite{Riis:2010zm,Formaggio:2011jg,SejersenRiis:2011sj,Esmaili:2012vg,Mertens-TAUP2015}.
Figure~\ref{fig:fut-see}
shows the KATRIN sensitivity presented in Ref.~\cite{Mertens-TAUP2015}.
One can see that it covers a significant portion of the
PrGLO allowed region.
Hence, there is a concrete possibility that
KATRIN can observe the effect of
$m_{4}$
if $\nu_{4}$ exists
and both
$m_{4}$ and $|U_{e4}|^2$ are not too small.

\subsection{Neutrinoless double-$\beta$ decay}
\label{sub:bb}

The implications
of non-standard mainly sterile massive neutrinos at the eV scale
for neutrinoless double-$\beta$ decay experiments
have been studied by several authors
\cite{Goswami:2005ng,Goswami:2007kv,Barry:2011wb,Li:2011ss,Rodejohann:2012xd,Giunti:2012tn,Girardi:2013zra,Pascoli:2013fiz,Meroni:2014tba,Abada:2014nwa,Giunti:2015kza,Pas:2015eia}.

\begin{figure*}[t]
\null
\hfill
\includegraphics*[width=0.49\linewidth]{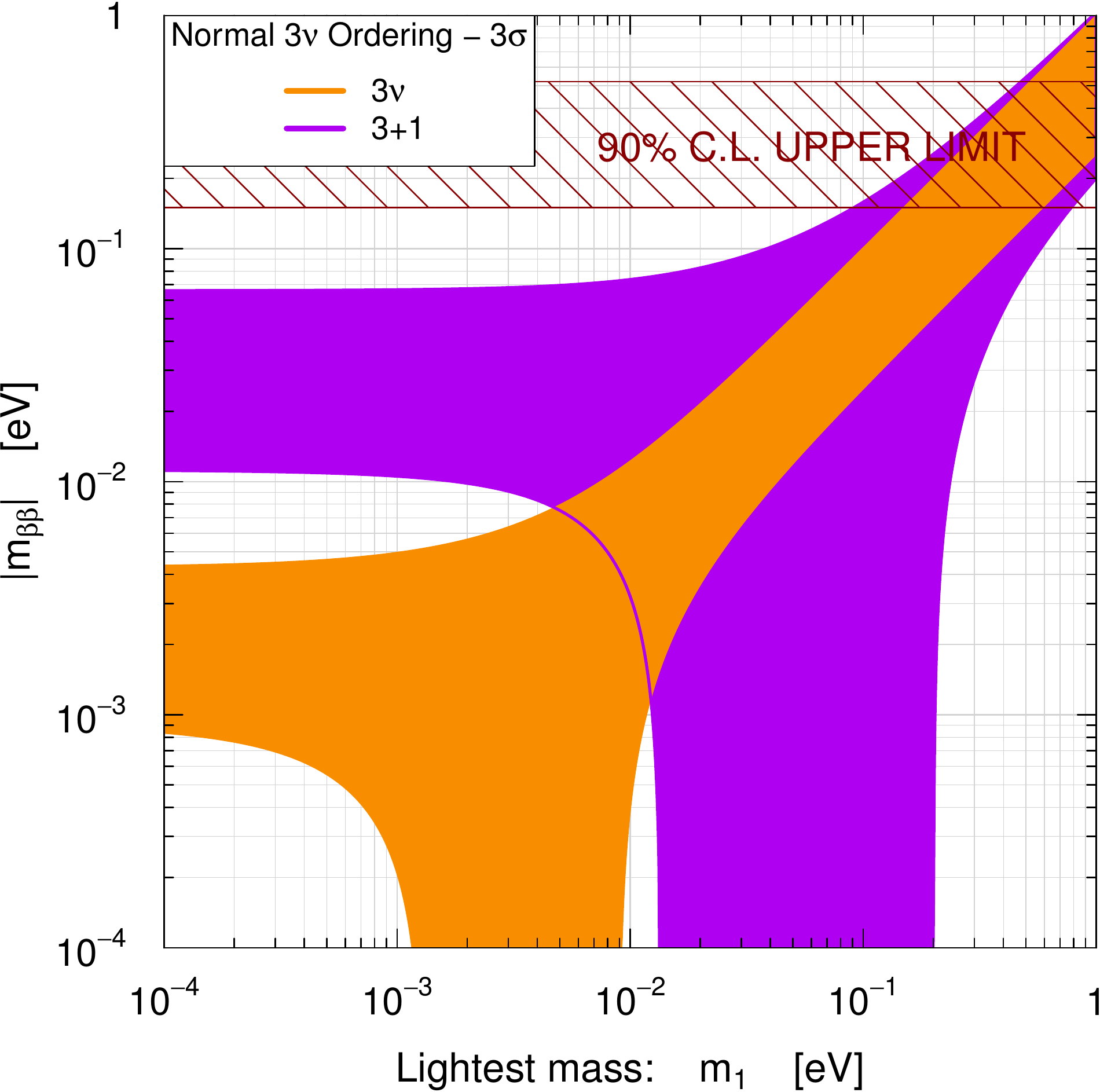}
\hfill
\includegraphics*[width=0.49\linewidth]{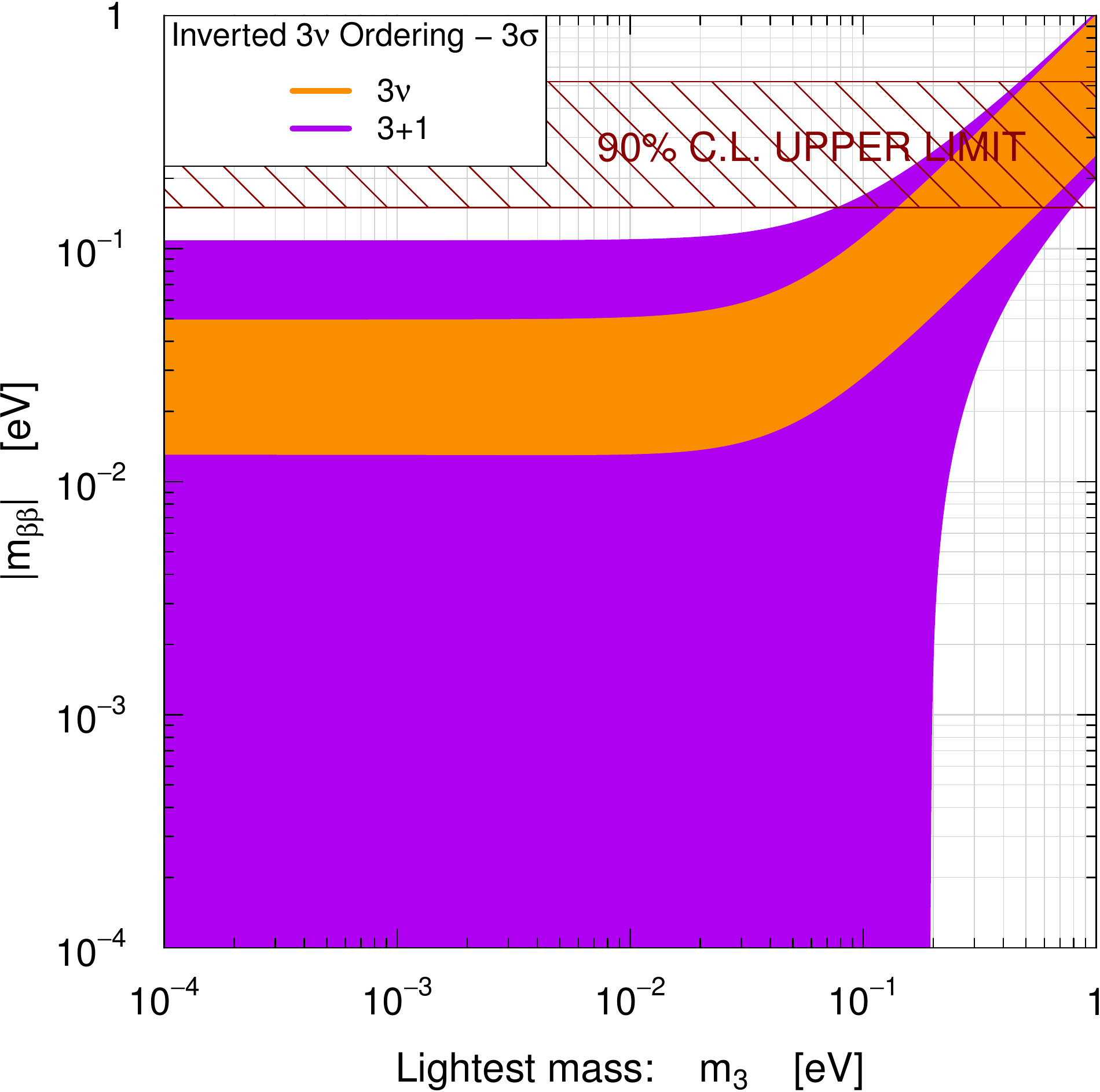}
\hfill
\null
\caption{ \label{fig:mbb}
Value of the effective Majorana mass $|m_{\beta\beta}|$
as a function of the lightest neutrino mass
in the cases of $3\nu$ and 3+1 mixing with normal and inverted ordering
of the three lightest neutrinos \cite{Giunti:2015kza}.
The horizontal band is an estimate of the current experimental
90\% C.L. upper limit for $|m_{\beta\beta}|$
taking into account the uncertainties of the nuclear matrix element calculations
\protect\cite{Bilenky:2014uka}.
}
\end{figure*}

If massive neutrinos are Majorana particles
(see the recent reviews in Refs.~\cite{Bilenky:2014uka,Pas:2015eia}),
in the case of 3+1 mixing
the rate of neutrinoless double-$\beta$ decay
is proportional to the square of the effective Majorana mass
\begin{align}
|m_{\beta\beta}|
=
\null & \null
\left|
|U_{e1}|^2 \, m_{1}
+
|U_{e2}|^2 \, e^{i\alpha_{2}} \, m_{2}
+
|U_{e3}|^2 \, e^{i\alpha_{3}} \, m_{3}
\right.
\nonumber
\\
\null & \null
\left.
+
|U_{e4}|^2 \, e^{i\alpha_{4}} \, m_{4}
\right|
.
\label{mbb}
\end{align}
In this expression there are three completely unknown complex phases
$\alpha_{2}$, $\alpha_{3}$, $\alpha_{4}$
which depend on the Majorana phases in the neutrino mixing matrix.
These unknown complex phases
can generate cancellations between the different mass contributions.
Figure~\ref{fig:mbb}
shows the range of allowed values of $|m_{\beta\beta}|$
as a function of the lightest neutrino mass
in the cases of $3\nu$ and 3+1 mixing with normal and inverted ordering
of the three lightest neutrinos \cite{Giunti:2015kza}.
The $3\nu$ mixing parameters are those obtained in Ref.~\cite{Capozzi:2013csa}
and the sterile neutrino mixing is that obtained in the global pragmatic 3+1 PrGLO fit
of short-baseline neutrino oscillation data
discussed in Section~\ref{sec:global}.

From Fig.~\ref{fig:mbb}
one can see that the presence of an additional
massive neutrinos at the eV scale
can change dramatically
the predictions for the possible range of values
of $|m_{\beta\beta}|$
\cite{Goswami:2005ng,Goswami:2007kv,Barry:2011wb,Li:2011ss,Rodejohann:2012xd,Giunti:2012tn,Girardi:2013zra,Pascoli:2013fiz,Meroni:2014tba,Abada:2014nwa,Giunti:2015kza,Pas:2015eia}.
In the case of a normal $3\nu$ mass hierarchy
($m_{1} \ll m_{2} \ll m_{3}$)
the value of
$|m_{\beta\beta}|$
is dominated by the contribution of $\nu_{4}$,
which implies that
$1 \times 10^{-2} \lesssim |m_{\beta\beta}| \lesssim 7 \times 10^{-2} \, \text{eV}$.
This range of values of $|m_{\beta\beta}|$ is
larger than that predicted by the standard $3\nu$ mixing in the case of a
normal hierarchy
and similar to that
predicted in the case of an inverted hierarchy in the standard $3\nu$ mixing scheme.
On the other hand,
in the case of an inverted $3\nu$ mass ordering
there can be a complete cancellation between
the contribution of $\nu_{4}$ and those of the three standard light neutrinos,
leading to the disappearance of the
lower limit for $|m_{\beta\beta}|$ predicted by the
standard $3\nu$ mixing scheme.

The next generation of
neutrinoless double-beta decay experiments
(see Refs.~\cite{GomezCadenas:2011it,Giuliani:2012zu,Schwingenheuer:2012zs,Cremonesi:2013vla,Artusa:2014wnl,Gomez-Cadenas:2015twa})
is planned to explore
the range of $|m_{\beta\beta}|$
between about
$1 \times 10^{-2}$ and $5 \times 10^{-2} \, \text{eV}$
predicted by the standard $3\nu$ mixing in the case of an
inverted hierarchy.
They are not expected to reach the
range of $|m_{\beta\beta}|$
between about
$8 \times 10^{-4}$ and $5 \times 10^{-3} \, \text{eV}$
predicted by the standard $3\nu$ mixing in the case of a
normal hierarchy.
From Fig.~\ref{fig:mbb} it is clear that
the predictions are dramatically changed in the 3+1 neutrino mixing scheme
and a positive result in these experiments
is guaranteed in the case of a normal mass hierarchy,
whereas in the case of an inverted mass hierarchy the allowed
range of $|m_{\beta\beta}|$ goes from zero to about 0.1 eV.

\section{Conclusions}
\label{sec:conclusions}

The reactor, Gallium and LSND anomalies can be explained by neutrino oscillations
if the standard three-neutrino mixing paradigm
is extended with the addition of light sterile neutrinos
which can give us important information on the new physics beyond the Standard Model.

The global fits of short-baseline neutrino oscillation data
in the framework of mixing schemes with one or more sterile neutrinos
suffer from a tension between the results of appearance and disappearance
short-baseline neutrino oscillation experiments.
This tension can be alleviated adopting the ``pragmatic approach'' advocated in Ref.~\cite{Giunti:2013aea},
in which the anomalous MiniBooNE low-energy excess of $\nu_{e}$-like events is neglected
from the global analysis of short-baseline neutrino oscillation data.
The cause of the MiniBooNE low-energy excess
is going to be investigated in the MicroBooNE experiment at Fermilab
\cite{Szelc:2015dga}.

Moreover,
the cosmological data indicate a tension between the necessity to have a sterile neutrino mass
at the eV scale and the expected full thermalization
of the sterile neutrinos through active-sterile oscillations in the early Universe
\cite{Archidiacono:2012ri,Archidiacono:2013xxa,Gariazzo:2013gua,Archidiacono:2014apa,Bergstrom:2014fqa}.
Hence,
the possible existence of light sterile neutrinos
at the eV scale
is controversial and needs new reliable experimental checks.

The impressive program of new experiments reviewed in Section~\ref{sec:perspectives}
gives us confidence that the question of the existence of the light sterile neutrinos
indicated by the reactor, Gallium and LSND anomalies
will be answered in a definitive way in the next years.

For neutrino physics,
the discovery of the existence of light sterile neutrinos would open a rich field of experimental and theoretical research
on the properties of the sterile neutrinos, their mixing with the active neutrinos and their
role in
neutrino experiments
(e.g. in solar
\cite{Dooling:1999sg,Giunti:2000wt,Giunti:2009xz,Palazzo:2011rj,Palazzo:2012yf,Giunti:2012tn,Palazzo:2013me,Long:2013hwa,Long:2013ota,Kopp:2013vaa},
long-baseline
\cite{deGouvea:2014aoa,Klop:2014ima,Berryman:2015nua,Gandhi:2015xza,Palazzo:2015gja},
and atmospheric
\cite{Goswami:1995yq,Bilenky:1999ny,Maltoni:2002ni,Choubey:2007ji,Razzaque:2011ab,Razzaque:2012tp,Gandhi:2011jg,Esmaili:2012nz,Esmaili:2013cja,Esmaili:2013vza,Rajpoot:2013dha}
neutrino experiments)
in astrophysics
(e.g. in supernova neutrino experiments
\cite{Caldwell:1999zk,Peres:2000ic,Sorel:2001jn,Choubey:2006aq,Choubey:2007ga,Tamborra:2011is,Wu:2013gxa,Esmaili:2014gya,Warren:2014qza}
and indirect dark matter detection
\cite{Esmaili:2012ut}),
high-energy cosmic neutrinos
\cite{Cirelli:2004cz,Donini:2008xn,Barry:2010en},
and in cosmology
(see Refs.~\cite{Lesgourgues-Mangano-Miele-Pastor-2013,RiemerSorensen:2013ih,Archidiacono:2013fha,Lesgourgues:2014zoa,Gariazzo:2015rra}).

Let us finally emphasize that the discovery of the existence of sterile neutrinos would be a major discovery
which would have a profound impact not only on neutrino physics,
but on our whole view of fundamental physics,
because sterile neutrinos are elementary particles beyond the Standard Model.
The existence of light sterile neutrinos would prove that there is new physics beyond the Standard Model at low-energies
and their properties can give important information on this new physics.
Without any doubt, such a discovery would deserve a new Nobel Prize in Physics.

\section*{Acknowledgments}
This work
was partially supported by the research grant {\sl Theoretical Astroparticle Physics} number 2012CPPYP7 under the program PRIN 2012 funded by the Ministero dell'Istruzione, Universit\`a e della Ricerca (MIUR).


\end{document}